# In the footsteps of Ebenezer Porter Mason and his nebulae

Jeremy Shears, Carl Knight, Martin Lewis, Lee Macdonald, Stewart Moore and Jeff Young

**Abstract**

In 1839 Ebenezer Porter Mason (1819-1840) produced detailed drawings of the Omega Nebula (M17), the Trifid Nebula (M20) and the eastern part of the Veil Nebula (NGC 6992 and 6995). He used a 12-inch (30 cm) reflector that he and his friends had built at Yale College, which at the time was the largest telescope in the USA. The drawings were remarkable for their accuracy and for his adoption of a new technique for delineating gradients in nebulosity using isophotes, or lines of equal brightness. This paper reviews his life and his observations, comparing his results with those of the modern amateur astronomer.

## Introduction

At the dawn of the nineteenth century, interest in the nebulae was gaining momentum. Messier's final catalogue had been published in 1781 and William Herschel's great sweeps of the night sky from Slough were adding many new nebulae to the tally, which grew to about 2500 objects. Much of Herschel's work was focussed on classifying the different types of nebulous object and thus his drawings were intended to represent the overall structure of the objects rather than precise details and positions. William's son John Herschel continued the work and set off for the Cape in 1833 to extend his father's catalogue of nebulae to the southern hemisphere. His Cape drawings took on a different form from his earlier work and that of his father, becoming what O.W. Nasim has described as *descriptive maps*, which attempted to depict precisely and faithfully the variations in brightness within each nebula and accurately to place the stars in and around the nebula (1). This approach, which owed much to the techniques of cartography and surveying, and involved careful measurements at the telescope, was aimed at determining whether changes in the structure of the nebulae occur over time. The Nebular Hypothesis, of which William Herschel was a great proponent, proposed that stars are formed by contraction and condensation from a nebulous fluid. John Herschel was more sceptical – he needed evidence of change. Would it therefore be possible to see this process happening by comparing accurate drawings of nebulae made at different epochs?

At about the same time as John Herschel's development of descriptive maps, the same approach was adopted, quite independently, by a comparatively little known astronomer in the United States: Ebenezer Porter Mason (1819-1840). In a paper read at the American Philosophical Society in 1840, Mason presented detailed observations of several nebulae (M17, M20 and the eastern part of the Veil Nebula) that he had made the previous year, whilst at Yale College using a telescope that he and his friends had constructed. The observations themselves were published in 1841 (2), the year after his tragic death at the age of 21, under the title "*Observations on nebulae with a fourteen feet reflector, made by H.L. Smith and E.P. Mason, during the Year 1839"*. Mason's aim was clearly laid out in the introduction to his paper:

"'The main object of this paper is to inquire how far that minute accuracy which has achieved such signal discoveries in the allied department of 'the double stars,' may be introduced into the observation of the nebulae, by modes of examination and description more peculiarly adapted to this end than such as can be employed in general reviews of the heavens."

In essence, Mason wanted to move the study of the nebulae beyond merely description, cataloguing and classification to one based on precise measurements that could be analysed and compared, much as had occurred in other branches of astronomy and the physical sciences. The current paper reviews Mason's techniques and observations, made with a 12-inch (30 cm) speculum reflector which was at the time the largest telescope in the USA (3), and presents recent observations of the same objects with modern equipment of a similar size.

**A promising life cut short**

Details of Mason's remarkable yet short life have been published elsewhere, notably in a book by Mason's mentor at Yale, Professor Denison Olmsted (1791-1859; Figure 1), "*Life and Writings of Ebenezer Porter Mason: Interspersed with Hints to Parents and Instructors on the Training and Education of a Child of Genius*" (4), so only a summary will be presented here (5).

Mason was born on 7 December 1819 in the village of Washington, Connecticut, where his father was pastor of the Congregational church. His mother died when he was 3 and at the age of 8 he was sent to live with his aunt in Richmond, Virginia, who encouraged his interest in astronomy. In 1830 Mason rejoined his father on Nantucket Island, Massachusetts, with his academic talents and enquiring mind already well in evidence.

Mason entered Yale College in 1835 at the age of 16, where he met Olmsted and the two became close friends. Mason began making astronomical observations with a 6-inch (15 cm) reflector borrowed from a classmate as well as with Yale's 5-inch (13 cm) Dollond refractor (6). Soon afterwards, he and several friends, including Hamilton Lanphere Smith (1818-1903: Figure 2) set about making a series of Herschelian reflectors, manufacturing their own speculum primaries from scratch. Having gained experience in telescope making, in the summer of 1838, Mason, Smith and another friend, Francis Bradley (1815-1893), embarked on the construction a 12-inch Herschelian reflector with a focal length of 14-feet (3.6 m):

"A tolerably good metal was cast, after several failures, and the speculum was finally polished near the close of summer. Mr. Smith and Mr. Bradley shared the expenses attending the formation of the mirror and erection of the telescope, and divided the long labor of grinding the speculum, and I united with them in the less tedious task of giving the mirror its final polish and figure".

They devised a framework for mounting the telescope based on the design employed by John Ramage (1783-1835) for his 25-foot reflector at the Royal Observatory Greenwich (Figure 3). Although the construction of the telescope was a team effort, it was Mason who led the way in its use, culminating in the observations of nebulae which he made with Smith during the summer of 1839, and which are described later.

After graduating in 1839, Mason remained at Yale and became Olmsted's assistant, helping him to prepare a new edition of Olmsted's *Introduction to Astronomy* for publication. He also prepared a supplement for this book entitled *Practical Astronomy*, which was essentially a handbook for the use of a range of astronomical instruments and performing calculations, and which received much critical acclaim. However, dark clouds were gathering, for whilst still an undergraduate it began to become apparent that Mason had contracted tuberculosis. During the summer of 1840 his health began to deteriorate. Thinking that a spell in the open air might help, he joined a field survey of the disputed Maine-Canadian border. The work was hard and soon took its toll on Mason's health. Leaving the survey, he went to New York when he completed the corrections to the proofs of *Practical Astronomy*, before boarding a ship to Richmond to stay with his uncle and aunt. He died there 11 days after his arrival, on 26 December 1840, just three weeks after his 21st birthday.

Smith also graduated in 1839 and moved back home to Ohio City, where he set up the 12-inch reflector on a more sophisticated version of the Ramage mount, although it was never again to achieve the same success as it had with Mason at Yale. Its ultimate fate is unknown. Smith went on to become Professor of Natural Philosophy and Astronomy at Kenyon College, Ohio, and later at Hobart College in New York. He carried out research in chemistry and pioneered the tintype photographic process (7).

**Mason's observational technique and his isophotes**

Mason adopted the same approach for each of his nebula drawings. He started by laying out a "groundwork of stars", concentrating on the brighter ones, which he then extended by "a kind of triangulation [which] was carried out by eye to all the stars in the neighbourhood". This involved comparing distances and angles between stars in much the same way as surveyors would triangulate via terrestrial features to draw a map. He then used this groundwork of stars as a framework upon which to sketch in the extent of the nebula. The whole process took place over several nights for a particular nebula. He concentrated his efforts on a limited number of nebulae "in order that the utmost accuracy in the delineation of the peculiar features and minutiae of these nebulae, attainable by protracted scrutiny, might be aimed at", although it is clear from his note that he observed many other objects in a more desultory way. He generally used a magnification of x80 when sketching, but he sometimes used x220 to resolve finer details.

The novel and innovative aspect of Mason's technique that contributed to the accuracy of his drawings was his use of *isophotes:* lines of equal brightness or brightness contour maps. This method he "hit upon for the exact representation of nebulae, which has essentially contributed to the accuracy of the…delineations….It was first suggested by the method usually adopted for the representation of heights above the sea-level on geographical maps, by drawing curves which represent horizontal sections of a hill and valley at successive elevations". In practice this meant "that if lines be imagined in the field of view winding around through all those portions of a nebula which have exactly equal brightness, these lines, transferred to out chart of stars, will give a faithful representation of the nebula and its minutiae".

In the autumn of 1839 Mason undertook a series of micrometer measurements of the positions of the brighter field stars. This further increased the accuracy of his drawings. For this work he was given permission to use Yale's 5-inch Dollond refractor (Figure 4) which was equipped with a parallel wire micrometer also by Dollond. However, this was tedious work since the instrument was on an altazimuth mount, a design which is not ideally suited to the use of the micrometer since the micrometer needed to be reoriented with respect to the lines of RA and Dec for each measurement.

Ultimately Mason combined his sketches made at the eyepiece, his isophote diagrams and his micrometer measurements to produce the final drawings which appear in his paper (8), along with catalogues of the positions of the stars he had measured. These drawings are reproduced in Figures 5 to 7. Since Mason was observing with a Herschelian reflector his drawings have south to the top and east to the left. For this publication, we have taken the liberty of reversing the drawings about the north south axis, to be consistent with the standard convention of representing objects as observed visually with astronomical telescope, with south to the top and east to the right

**The Trifid Nebula, M20, NGC 6514**

Mason observed the Trifid Nebula in Sagittarius on six nights in July and August 1839 using the 12-inch reflector. On the first two nights he could only make out the well-known brighter (southern) emission component of the nebula, in which the “three clefts which divide it were made out without difficulty”. Then on Aug 7 he glimpsed the northern component, now known to be a reflection nebula, becoming one of the first people to record this section of the nebula: “Mr. Smith and myself both remarked that the large star …. immediately adjoining [the main part of the nebula] was surrounded with a distinct nebula, not far inferior in brightness to the trifid. It was scarcely to be overlooked, and was seen at the first glance into the field. Its limits are nearly as great as those of the trifid nebula, with which it is nearly, or quite, in contact”. Mason noted that according to his analysis of the records, neither William nor John Herschel had detected the reflection nebula.

Mason’s completed drawing of M20 and the isophote sketch upon which it was based are shown in Figure 5a and b. The modern drawings, made with a 25 cm reflector (Figure 5d) and a 30 cm SCT (Figure 5c) and UHC nebular filters (9), show more detail, with the latter showing considerably more, as well as nebulosity elsewhere in the field. Both clearly show the fainter emission nebula, which nowadays is frequently picked up with much smaller instruments.

**The Omega Nebula, M17, NGC 6618**

Mason observed the Omega Nebula in Sagittarius on 5 nights in August 1839. By the end of the fourth night, 19 August, he had placed about 30 stars in the field to his satisfaction. On 14 August he prepared an isophote sketch, but observations on a sixth night, 19 August, were aborted due to interference from moonlight. Mason commented that he would have liked to make further observations, “but such now is impractical”, presumably because by the time the moon had waned, Smith would have left Yale for home. The final drawing is shown in Figure 6a.

Modern observations with 22 to 46 cm instruments are shown in Figure 6b to c. They confirm Mason’s assertion that the western part of the loop is fainter than the eastern. Mason noted two prominent “knots” at the internal angle of the nebula where the loop joins the bright bar which is the most prominent part of the nebula, with the southern of the two being brighter and possibly being separate from the rest of the nebula. These features can also be seen in two of the modern sketches. Mason drew attention to “a slight protrusion from the upper curve of the bend" of the loop which involves a star – the feature is obvious in Figure 6d for which a UHC filter was employed. Something about which Mason was "nearly certain", and which was also seen by Smith, is a faint patch of nebulosity to the south of the bright bar – his drawing shows it as a faint curve, although he noted that the "extremely faint light" filled much of the space between this star and the main part of the nebula. This feature is shown in all the modern drawings, but is especially prominent in Figure 6d.

**The eastern Veil Nebula, NGC 6992 and 6995**

At the time Mason made his observations the term "Veil Nebula" (or Bridal Veil Nebula) had not been coined (10). Instead he refers to it as the "great Nebula Cygni". He examined it in detail on seven nights in August 1839, but he also made "several observations of little importance in July" that he did not record.

The initial discovery of parts of the eastern Veil were made by William Herschel, but were later observed in more detail by John Herschel. The latter recorded two separate nebulae in the region: h 2092 and 2093, although he had suspected a faint band of nebulosity between them. It was evident to Mason and Smith on their first night of detailed observations, 1 August, that they were in fact components of a single extended object: "Mr. Smith and myself were both able to trace the nebula continuously from one to the other, and the reverse, so that these are now satisfactorily ascertained to constitute one immense nebula, stretching through several fields of 30' diameter….the whole cannot be less than 2° or 3° long".

The sheer extent of the nebula and the number of stars in the field caused him difficulties in producing the "groundwork or stars" upon which he would make his drawing. To protect his night vision, he normally used a lamp the light of which he could extinguish whilst observing. However, in the case of the Veil this resulted in very slow progress. "I therefore fixed a lamp just beneath my feet, on the ladder of steps, and found I could, by this means, easily copy the stars from the field on a sheet of paper".

The similarity between Mason's drawing (Figure 7a) and those made with modern telescopes (Figure 7b, c and d), for which nebular filters were also employed, is noteworthy. At the southern end of the object is a complex region of nebulosity. Mason commented on it thus: "The peculiar characteristic of the upper [southern] portion of the nebula cannot be mistaken. I can think of no comparison as good as that which Herschel gives it, that of a network, or interlacing of nebulae". He therefore referred to this as "the network portion" of the object, to distinguish it from the northern part, which he called "the bifurcate", since the nebula appears to split into two at its northern end. Near the bifurcation Mason suspected the nebula "to break up into rifts and braches nearly parallel", something of which can be seen in the modern drawings.

**Mason's legacy and the modern amateur astronomer**

By the age of 19, Mason had already demonstrated himself to be a talented astronomer, with enormous potential. It is possible that had his life not been cut short he would have gone on to become a great astronomer. Indeed, one wonders what further discoveries and insights he would have achieved. His paper on the nebulae stands as testimony to his observation skills, to his innovative capacity to transfer methods from other sciences such as surveying and cartography to achieve great accuracy, and to his tenacity in overcoming the practical difficulties inherent in the basic equipment he had available to make precise measurements. The results were drawings of nebulae of unprecedented accuracy and it is not surprising that Mason's work received high praise from John Herschel: (11) "Mr. Mason, a young and ardent astronomer, a native of the United States of America, whose premature death is the more to be regretted, as he was (as far as I am aware) the only other recent observer who has given himself, with the assiduity which the subject requires, the exact delineations of nebulae, and whose figures I find at all satisfactory". John Herschel also corrected his position of part of the Veil Nebula which he had recorded in one of his catalogues when he saw Mason's measurements, noting that "the conclusion from Mr. Mason's obs. is irresistible".

A brief comparison of Mason's results obtained with his two instruments, a rude 12-inch telescope with a homemade speculum mirror and an unwieldy mount, and a 5-inch refractor on a simple altazimuth mount, although admittedly impressive instruments for the period, with results obtained with equipment available to today's amateur astronomers should remind us how fortunate we are. Clearly it is not possible to draw definitive conclusions about the relative performance of Mason's 12-inch telescope since there are too many other variables, such as local observing conditions and light pollution, elevation of the objects, observer experience, physiology and expectations, the other optics involved such as secondary mirrors (12) and eyepieces, etc. One big difference between modern observers and Mason is that nowadays we have access to nebular filters, which can improve the apparent contrast of an object making it easier to discern detail in the nebulae. These can dramatically change the visual appearance of emission objects with the Veil, for example, apparently becoming a totally different object when a filter is employed. Nevertheless our results with modern 8 to 10-inch telescopes certainly stand comparison with Mason's.

Much has also been written on the efficiency of speculum mirrors compared to aluminium on glass mirrors, but here too the variables are many, such as the chemical composition of the speculum, the fact that the speculum reflectivity rapidly deteriorates with time, accuracy of the reflecting surface, not to mention the fact that reflectivity of speculum and aluminium varies with wavelength. For example, in one study, Stephen James O'Meara considered the reflectivity of speculum and aluminised glass mirrors and concluded that William Herschel's "Large 20-Foot" telescope with an 18.7-inch (47 cm) mirror was roughly equivalent to a modern 10-inch reflector with excellent coatings (13).

Whatever the comparison, the point is moot and we should simply admire Mason's pioneering work, whilst seeking to use our own telescopes, whatever they might be, to the best of our ability, perhaps employing the methods that Ebenezer Porter Mason has handed down and striving to achieve the same accuracy as he did.

**Acknowledgements**

The authors thank Joanna Hopkins and Rupert Baker of the Royal Society library for scanning the drawing's from Mason's original papers. Given the delicacy of reproduction of the original publication, and their low contrast, this was a real challenge. Annette van Aken of the Peabody Museum at Cambridge, Massachusetts, obtained the photographs of the Yale 5-inch telescope used by Mason, which is in the care of the Museum, and gave permission to reproduce them in this paper. David Simkin provided the photograph of Hamilton Lanphere Smith.

JS is indebted to Richard Baum for presenting him with a copy of Mason's paper on the nebulae, which stimulated his interest in the subject, and for his great encouragement to pursue this research.

**Addresses**

JS: 'Pemberton', School Lane, Bunbury, Tarporley, Cheshire, CW6 9NR, UK
[bunburyobservatory@hotmail.com]

CK: Kilkern Road, RD1, Bulls 4894, New Zealand [Observatory@ngileah.co.nz]

ML: 21 Hazelwood Drive, St Albans, Hertfordshire, AL4 0UP, UK
[martin@skyinspector.co.uk]

LM: c/o BAA, Burlington House, Piccadilly, London' W1J 0DU, UK
[leetmacdonald@gmail.com]

SM: Conifers, New Town Road, Thorpe-le-Soken, Essex, CO16 0ER, UK[sigarro@btinternet.com]

JY: Rokeby Hall, Grangebellew, Co. Louth, Ireland [jeff@rokeby.ie]

**Notes and references**

1. Nasim O.W., Studies in the History and Philosophy of Science, 42, 67-84 (2001). Nasim argues for a continuity in purpose between the "descriptive maps" of John Herschel and E.P. Mason and the first photographs of the nebulae made in the 1880s.

2. Mason E.P., Transactions of the American Philosophical Society, New Series, 17, 165-213 (1841).

3. According to the late Joseph Ashbrook of Sky & Telescope, the largest at the time was probably an 8½-inch (22 cm) reflector built by Amasa Holcomb (1787-1875) of Southwick, Massachusetts.

4. Olmsted D., Life and Writings of Ebenezer Porter Mason, publ. Dayton and Newman (1842). The book can be read online in various places including Google Books.

5. Other biographies are mainly based on material in Olmsted's book. E.g. "Notes on a Forgotten Episode": Treadwell T.R., Popular Astronomy, 51, 497-500 (1942). Joseph Ashbrook also prepared a summary of Mason's life for the December 1972 edition of Sky & Telescope, page 366. This was reproduced in: Ashbrook J., The Astronomical Scrapbook, publ. CUP (1984), Chapter 74, "E.P. Mason and the nebulae".

6. The 5-inch Dollond had been acquired by Yale in 1829. It was rather poor mounted on an altazimuth mount. Castors enabled it to be moved from window to window in the Athaneum at Yale, which did for an observatory. Nevertheless in August 1835 Denison Olmsted and Elias Loomis used it to detect Halley's Comet - the first recovery observation from the USA of that particular apparition.

7. This led to the popularisation of photography in America. He patented the tintype invention on 19 February 1856. In 1848 Smith wrote "The World", one of the first science textbooks written in America.

8. Smith also observed the objects, although it appears that the drawings were made by Mason from Mason's own observations. Mason noted: "I regret that I have at hand no notes of Mr. Smith's observations distinct from my own, and can, therefore, furnish only such scattered remarks of his as I happened to record at the time. This deficiency is of no real consequence, since it was the constant practice for each of us to verify the observations of the other".

9. A UHC filter, or Ultra-High Contrast nebular filter, is often used by modern deep-sky observers to bring out detail in emission nebulae. The typical transmission range is from 484 to 506 nm. It transmits both the O-III and H-beta spectral lines.

10. The Veil Nebula was discovered by William Herschel in 1784 - the part now known as NGC 6992. John Herschel discovered the other component discussed here, NGC 6995 in 1825. William also discovered the western part of the Veil in 1784, NGC6960.

11. Herschel J., Results of Astronomical Observations Made During the Years 1834, 5, 6, 7, 8 at the Cape of Good Hope. Publ. Smith, Elder & Co., London (1847).

12. Mason's 12-inch telescope, being a Herschelian design, had no secondary mirror.

13. O'Meara S.J., Steve O'Meara's Herschel 400 Observing Guide, publ. CUP (2007).

14. The original image is in Special Collections section of Kenyon College Archives.

15. This artwork is from 'An Introduction to Practical Astronomy', published by William Pearson in several volumes between 1824 and 1829.

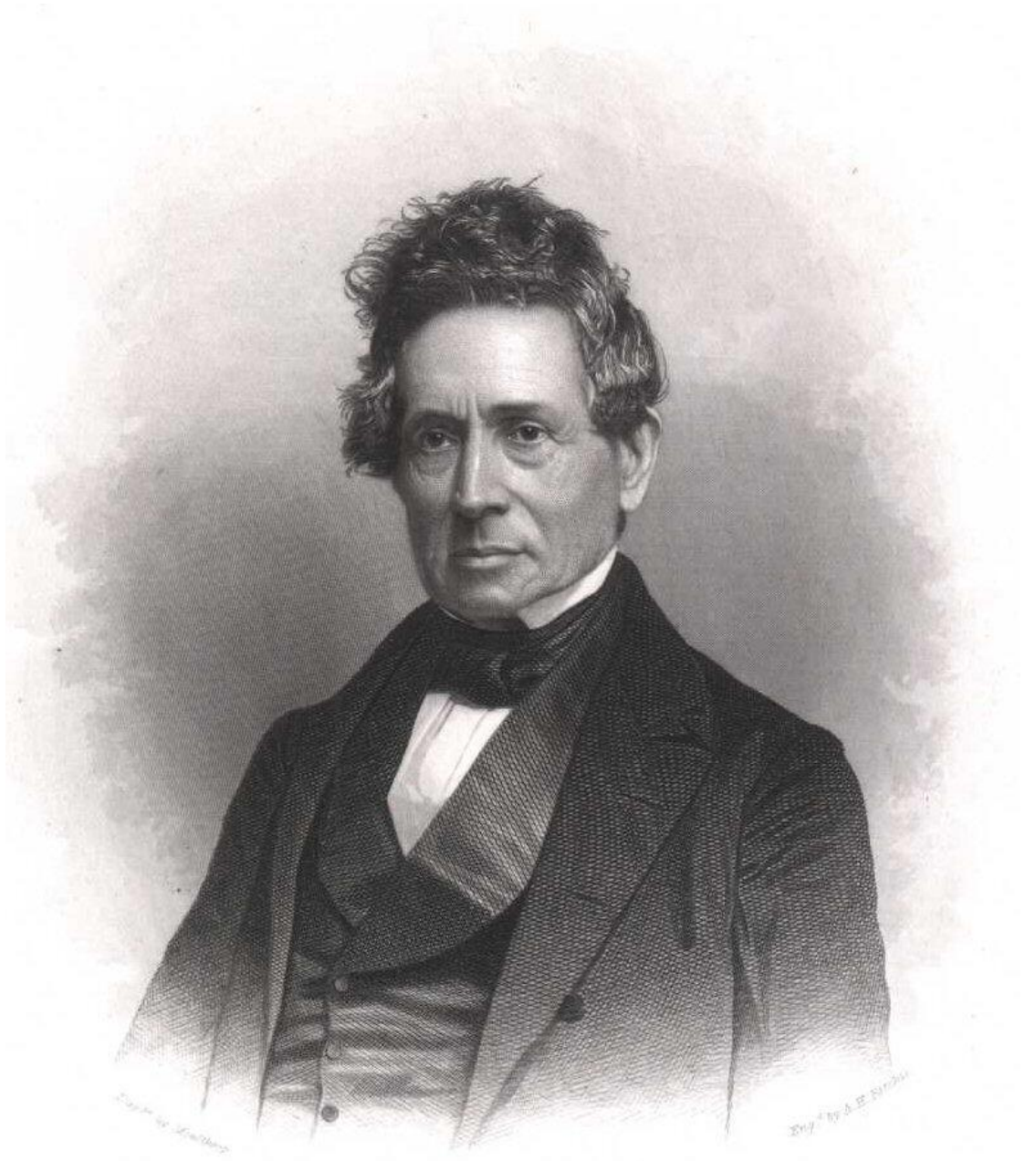

Figure 1: Denison Olmsted (1791-1859)

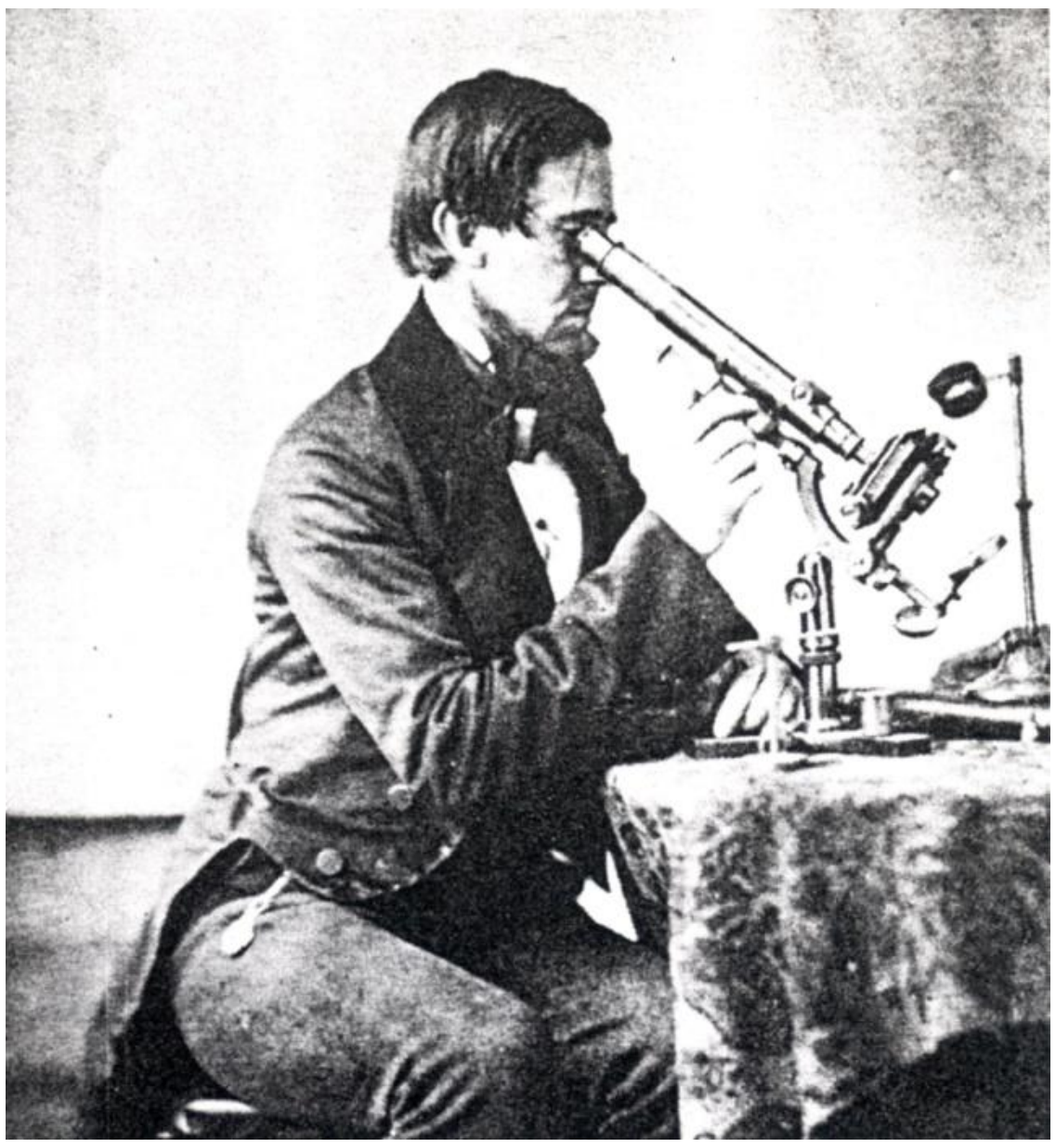

Figure 2: Hamilton Lanphere Smith (1818-1903)

(image courtesy of David Simkin (14))

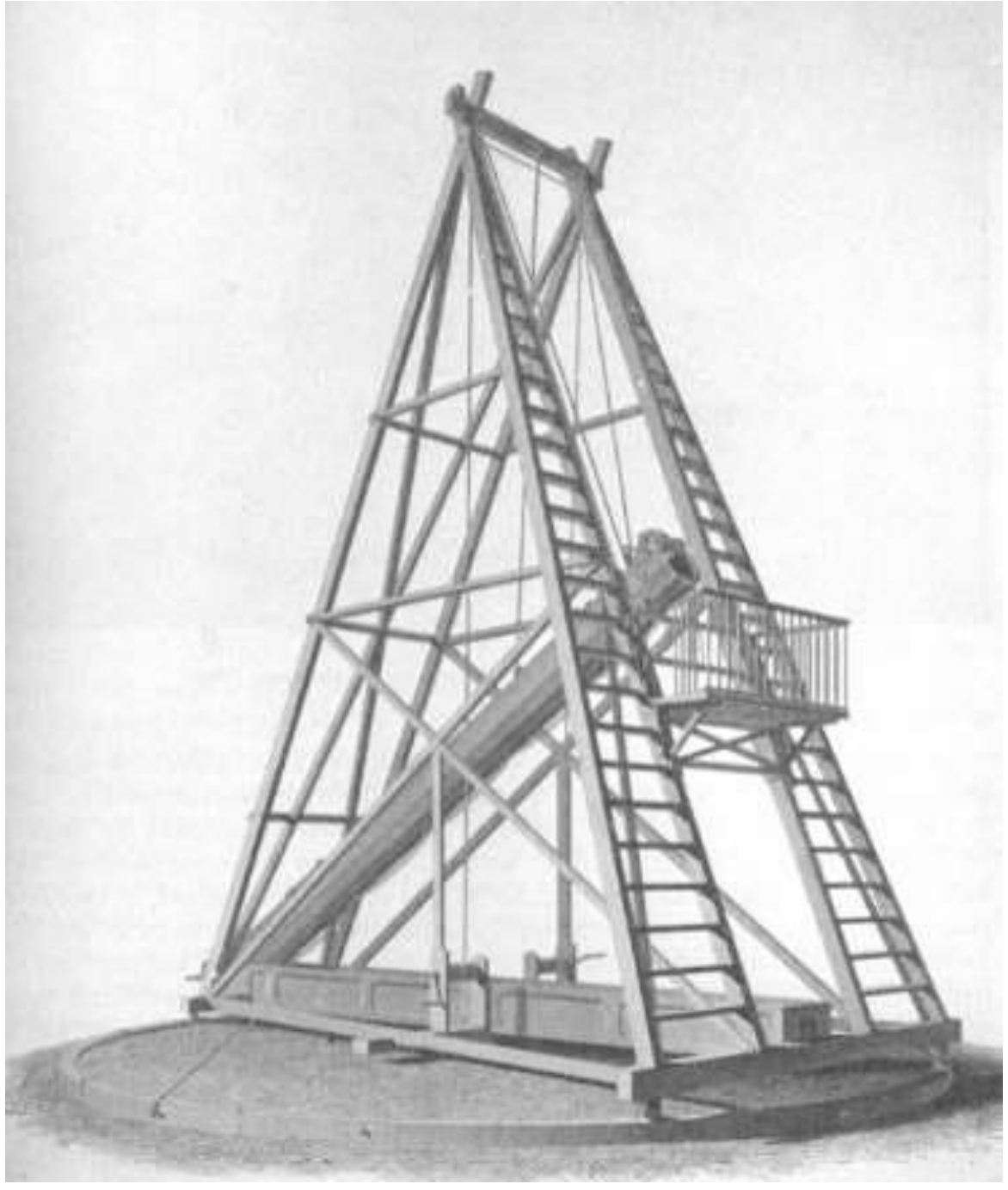

Figure 3: Ramage's 25-foot reflector erected in 1820 at Greenwich upon which the design of the mount for Mason’s 12-inch telescope was based

(RAS archives (15))

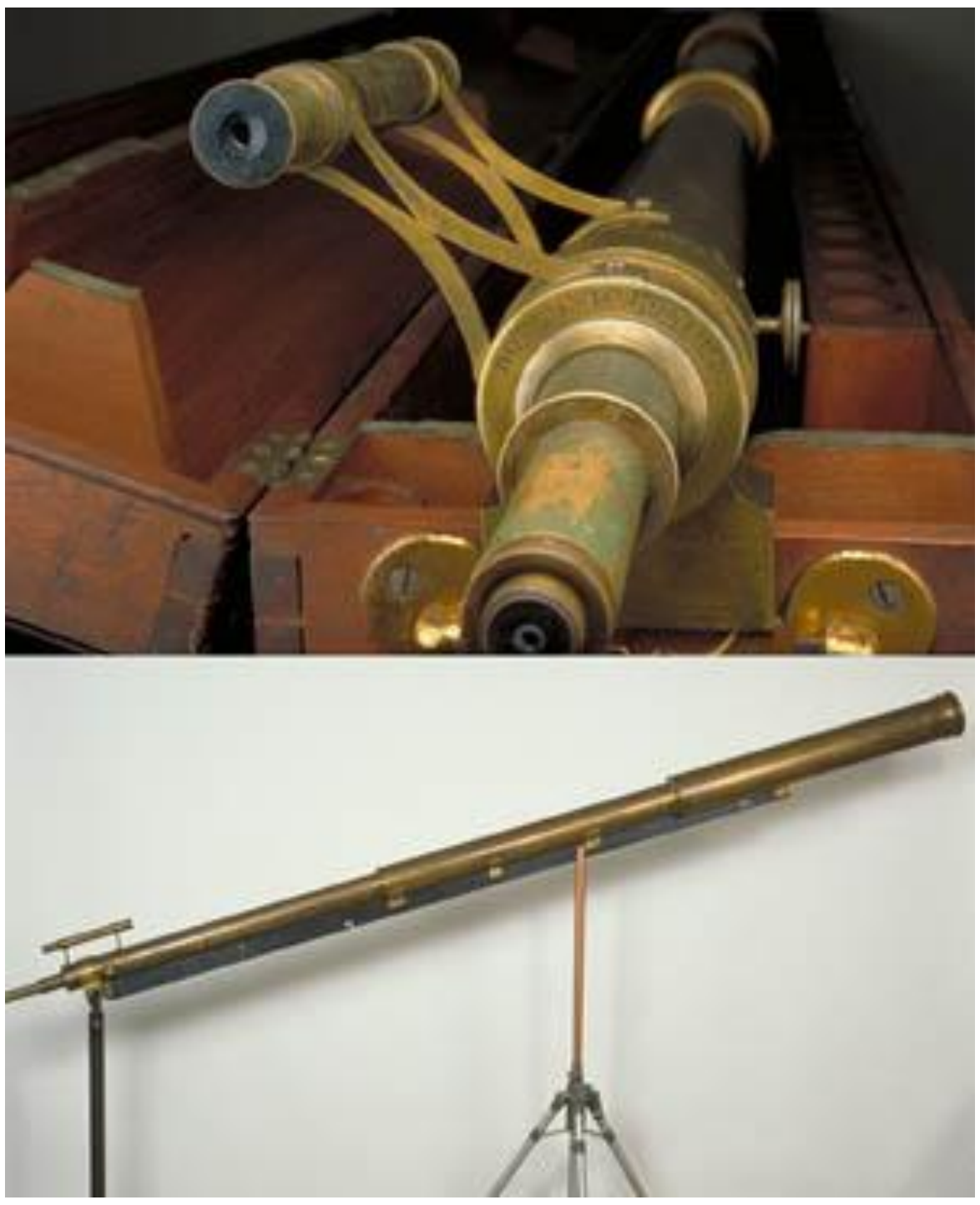

Figure 4: the 5-inch Dollond refractor used by Mason at Yale

(image courtesy of the Peabody Museum, Cambridge, Massachusetts)

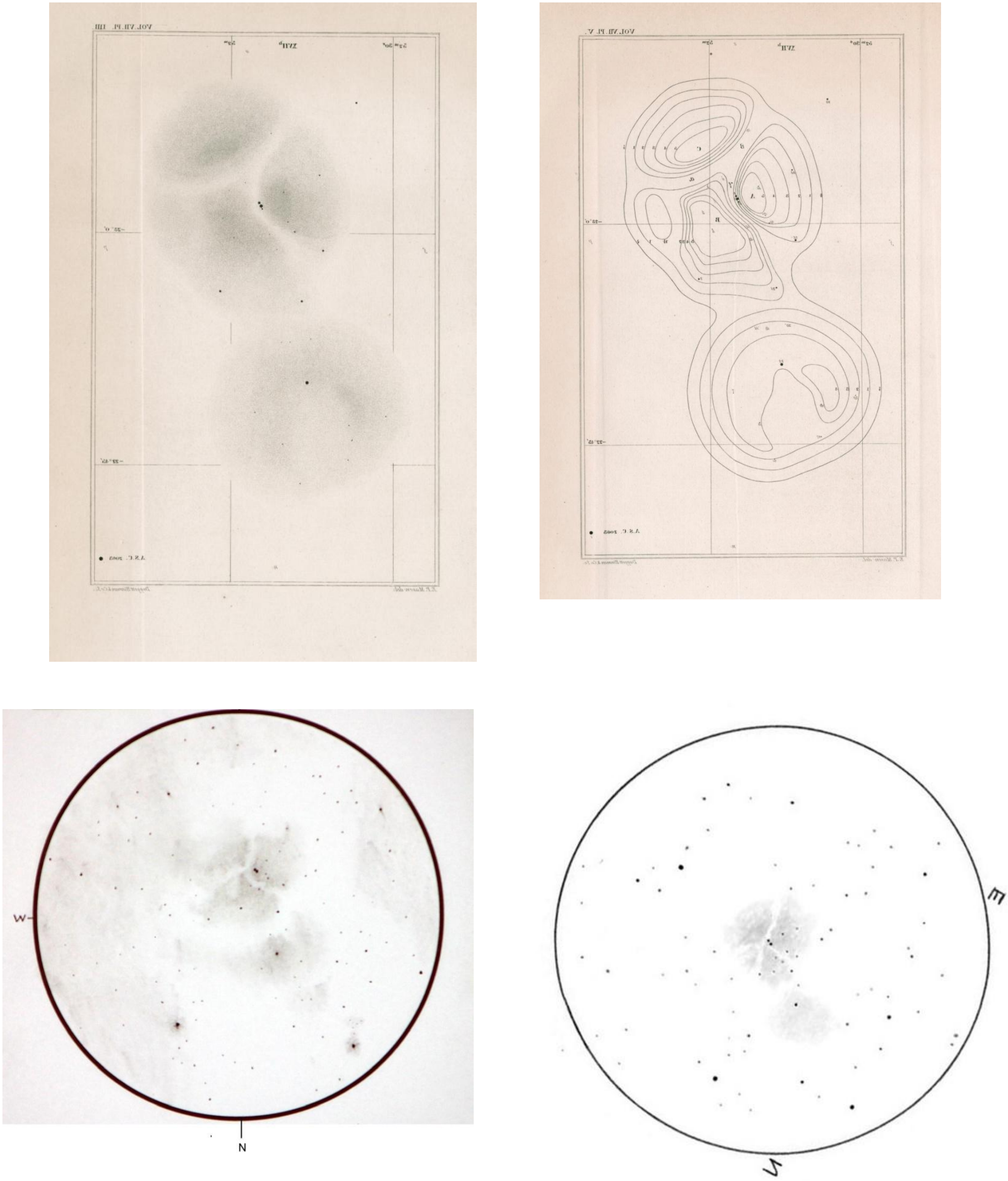

Figure 5: The Trifid Nebula, M20. (a) Top left –E.P. Mason final drawing; (b) Top right: E.P Mason isophote drawing; (c) Bottom left: Carl Knight, 3 May 2013, 30 cm SCT, UHC filter, x80, field 54'; (d) Bottom right: Jeff Young, 23 July 2009, 25 cm Cass, UHC filter, x100

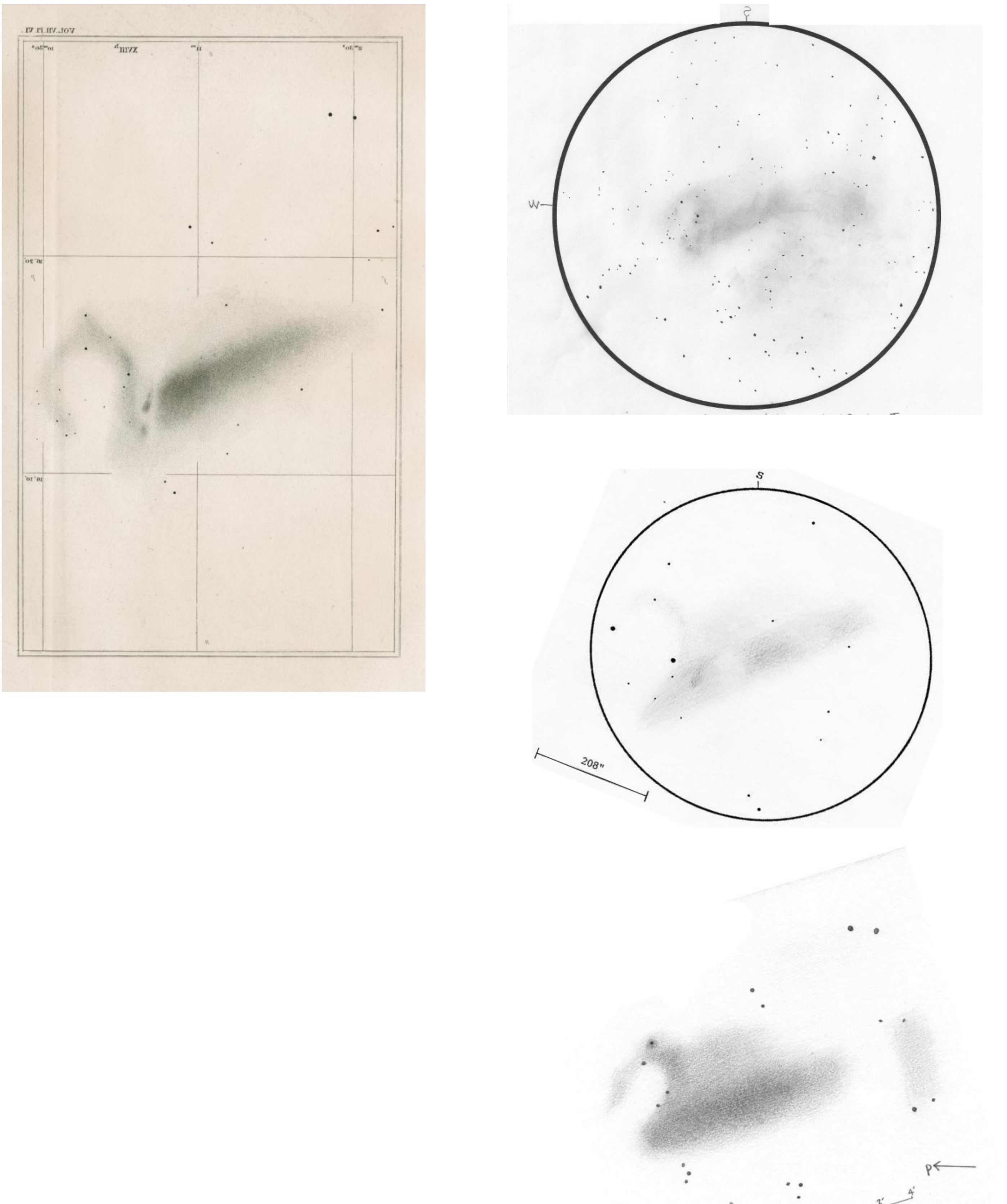

Figure 6: The Omega Nebula, M17. (a) Left –E.P. Mason final drawing; (b) Top right: Carl Knight, 23 September 2011, 30 cm SCT, x80, field 54’; (c) Middle right: Lee Macdonald, 12 July 1993, 22 cm Newtonian, x90; (d) Bottom right: Martin Lewis, 2 October 2005, 46 cm Newtonian, x156, UHC filter

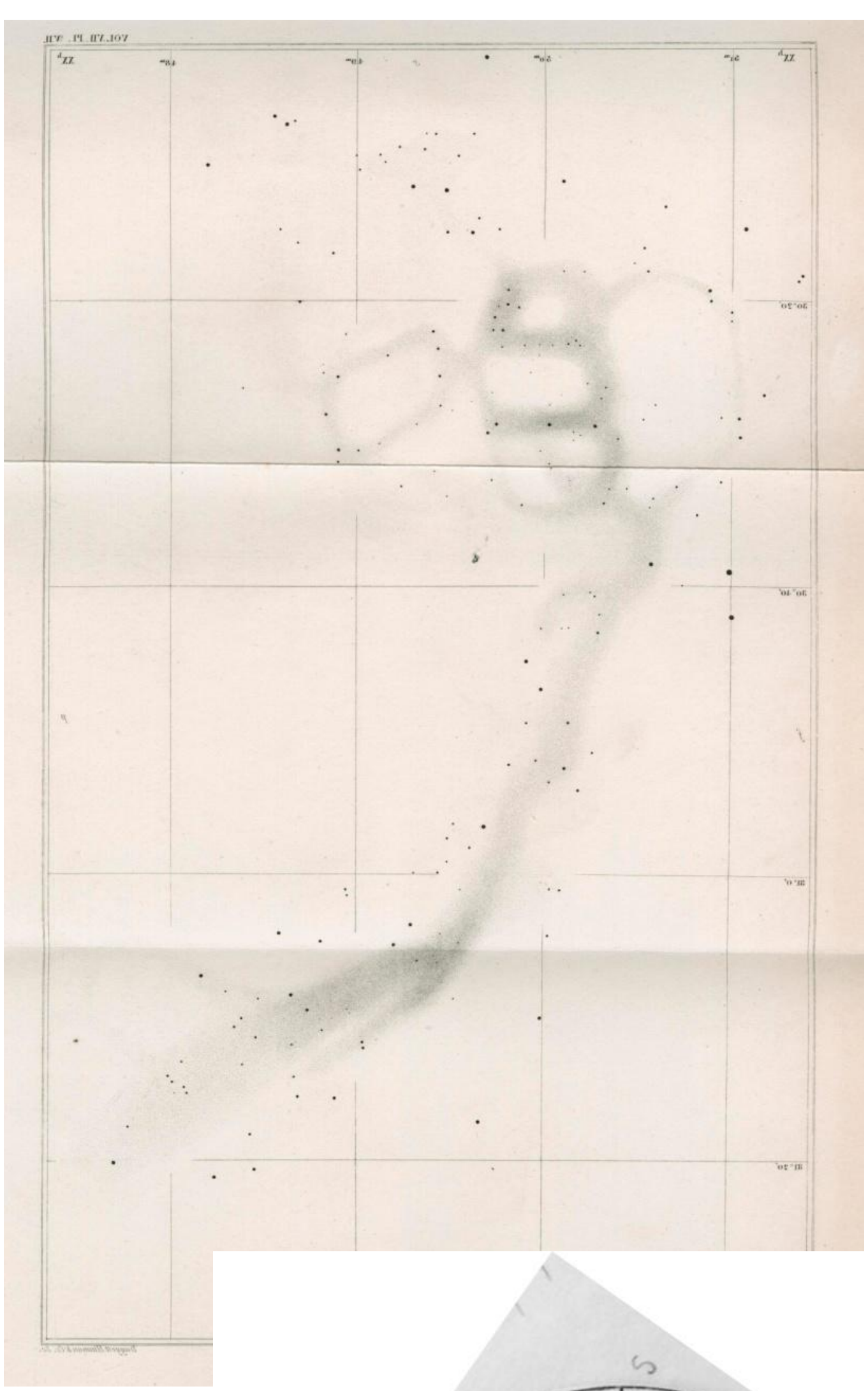

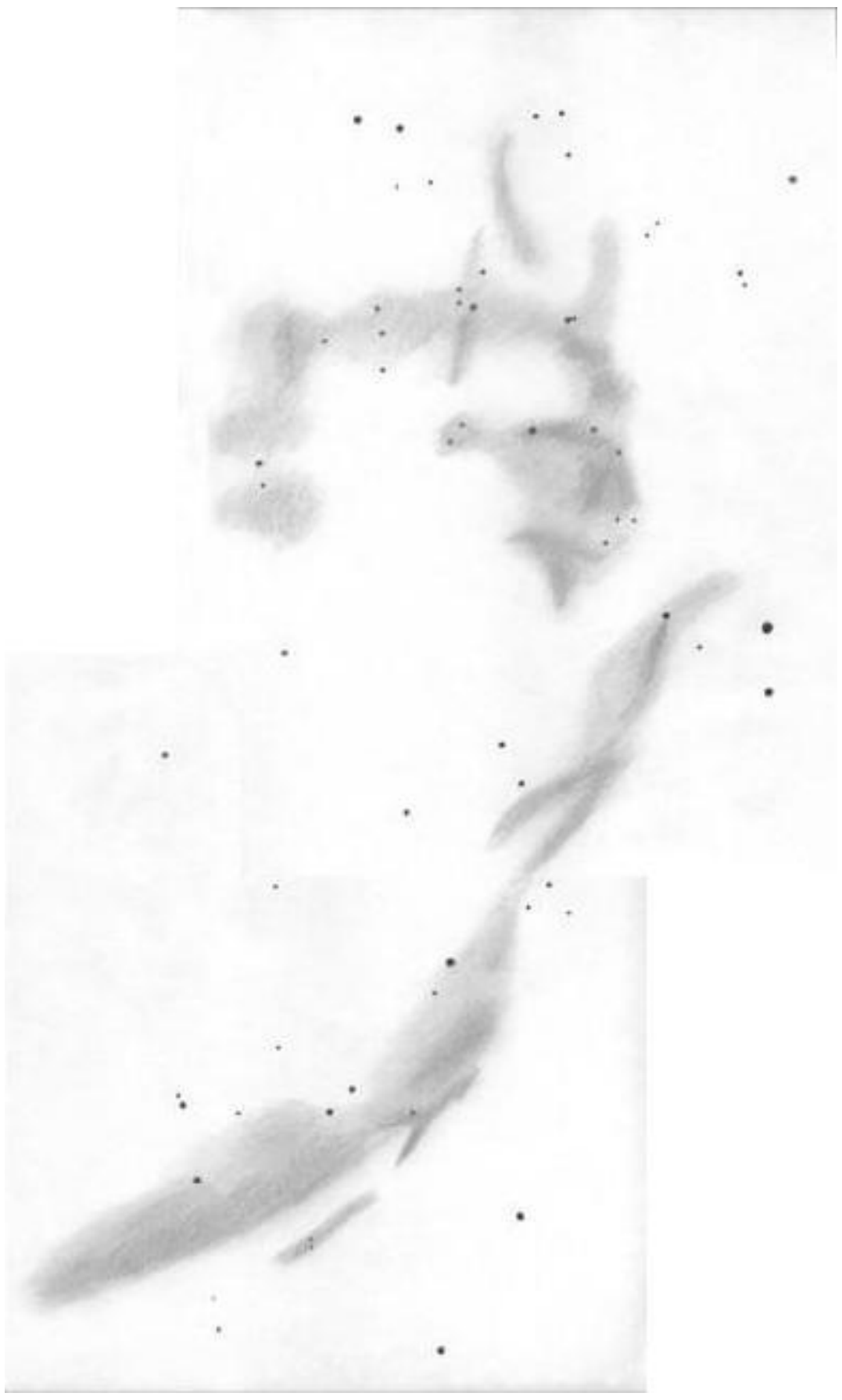

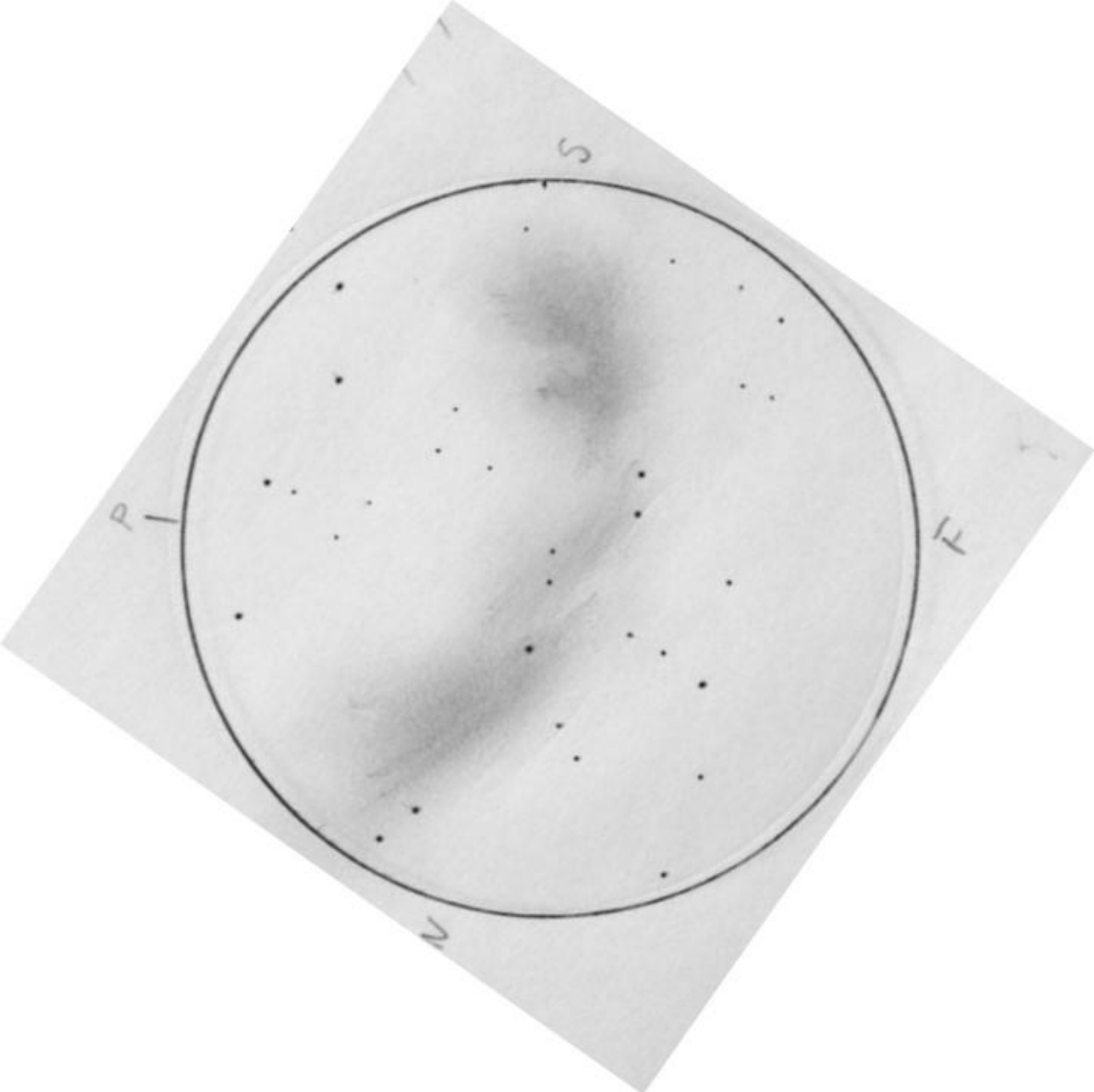

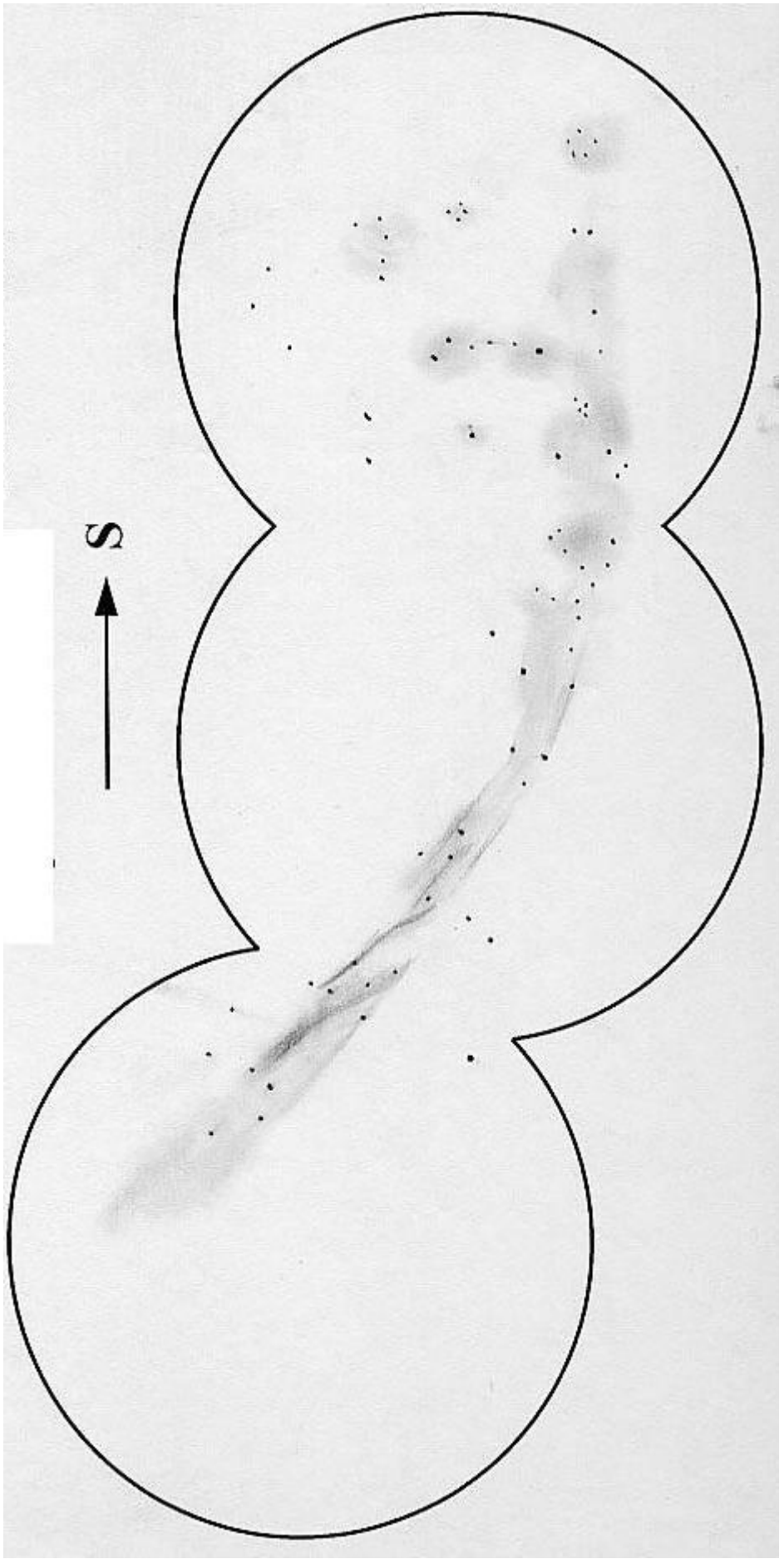

Figure 7: The eastern Veil Nebula, NGC 6992 and 6995. (a) Top left –E.P. Mason final drawing; (b) Top right: Martin Lewis, 8 & 20 September 2009, 46 cm Newtonian, x101 and x 127, O-III filter; (c) Bottom right: Stewart Moore, 36 cm Newtonian, x100, UHC filter, each of the three fields covers 30’, (d) Bottom left, Carl Knight 30 cm SCT, x80, UHC filter, field 54’, 2006 November 11

·itish